\documentclass[draft]{agujournal2019}
\usepackage{url} 
\usepackage{lineno}
\usepackage[inline]{trackchanges} 
\usepackage{soul}

\draftfalse

\journalname{Space Weather}

\begin{document}

\title{Radiation Data Portal: Integration of Radiation Measurements at the Aviation Altitudes and Solar-Terrestrial Environment Observations}

\authors{V.M. Sadykov\affil{1,2}, I.N. Kitiashvili\affil{1}, W. K. Tobiska\affil{3}, M. Guhathakurta\affil{1,4}}
\affiliation{1}{NASA Ames Research Center, Moffett Field, CA 94035, USA}
\affiliation{2}{Bay Area Environmental Research Institute, Moffett Field, CA 94035, USA}
\affiliation{3}{Space Environment Technologies, Pacific Palisades, CA 90272, USA}
\affiliation{4}{NASA Headquarters, Washington DC 20546, USA}

\correspondingauthor{Viacheslav Sadykov}{sadykov@baeri.org}

\begin{keypoints}
\item Radiation Data Portal is a web application for search and exploration of in-flight radiation measurements.
\item The portal integrates the radiation dose rates obtained with the ARMAS device and GOES soft X-ray and proton fluxes.
\item The impact of solar proton events (SPEs) on the ARMAS radiation dose measurements at aviation altitudes is discussed.
\end{keypoints}

\begin{abstract}
The impact of radiation dramatically increases at high altitudes in the Earth’s atmosphere and in space. Therefore, monitoring and access to radiation environment measurements are critical for estimating the radiation exposure risks of aircraft and spacecraft crews and the impact of space weather disturbances on electronics. Addressing these needs requires convenient access to multi-source radiation environment data and enhancement of visualization and search capabilities. The Radiation Data Portal represents an interactive web-based application for search and visualization of in-flight radiation measurements. The Portal enhances the exploration capabilities of various properties of the radiation environment and provides measurements of dose rates along with information on space weather-related conditions. The Radiation Data Portal back-end is a MySQL relational database that contains the radiation measurements obtained from the Automated Radiation Measurements for Aerospace Safety (ARMAS) device and the soft X-ray and proton flux measurements from the Geostationary Operational Environmental Satellite (GOES). The implemented Application Programming Interface (API) and Python routines allow a user to retrieve the database records without interaction with the web interface. As a use case of the Radiation Data Portal, we examine ARMAS measurements during an enhancement of the Solar Proton (SP) fluxes, known as Solar Proton Events (SPEs), and compare them to measurements during SP-quiet periods.
\end{abstract}

\section*{Plain Language Summary}
The Radiation Data Portal is an interactive web-based application for convenient search, visualization, retrieval, and exploration of in-flight radiation measurements and conditions of the solar-terrestrial environment. The database integrates radiation measurements obtained with the ARMAS device at aviation altitudes and the soft X-ray and proton fluxes measured by the GOES satellites. The radiation measurements are analyzed for flights during a solar proton flux enhancement, a so-called solar proton event, and compared to measurements when the level of the proton flux is low.

\section{Introduction}
\label{section:introduction}

Variations of the Earth's radiation environment and Space Weather disturbances are a subject of primary importance from scientific, operational, and commercial points of view. Understanding the present state of the radiation environment at aviation altitudes, its connection to solar transient activity (e.g., precipitation of solar energetic particles), and extra-solar radiation (galactic cosmic rays) is critical for estimation of potential safety and technological risks. To address the problem of efficient utilization of available radiation environment data, we developed the Radiation Data Portal, which integrates measurements from different sources, enhances data accessibility, and provides a search engine supported by visualization tools for efficient identification and overview of the data of interest.

In-situ measurements during airplane flights and balloon experiments represent the most direct way to get information about radiation environment conditions. The Automated Radiation Measurements for Aerospace Safety \cite<ARMAS,>{Tobiska15,Tobiska16,Tobiska18} program obtains measurements of the local radiation environment conditions, dose, and dose rates from dosimeters flown on a commercial aircraft. The data are retrieved in real time, downlinked to the ground, and compared against the Nowcast of the Atmospheric Ionizing Radiation for Aerospace Safety \cite<NAIRAS v1,>{Mertens13} radiation environment model.

Another example of a radiation measurement campaign is the routine measurements made with a Liulin silicon spectrometer organized into a database \cite{Ploc13}. Liulin-type of spectrometers are widely used for radiation absorption dose measurements in the space radiation environment \cite{Dachev15}. The database currently consists of measurements of the radiation doses and energy deposition spectra since 2001. It contains over 10$^5$ individual measurements. In addition to airplane in-flight experiments, high-altitude balloon experiments provide an essential grounding for detector cross-comparison \cite{Straume16} and model verification \cite{Joyce16}.

The radiation environment is primarily determined by solar-terrestrial interactions and galactic cosmic rays. Therefore, integration of the absorption-dose measurements and observations that characterize space weather conditions are necessary for understanding the radiation environment. The growing number of measurements requires the development of tools to perform efficient analysis of data from multiple sources. The critical needs of efficient data search, retrieval, and analysis encourage development of multi-datasource web platforms. For example, this approach has been implemented to analyze multi-instrument observations of solar flares, leading to the Interactive Multi-Instrument Database of Solar Flares (IMDSF, \url{https://data.nas.nasa.gov/helio/portals/solarflares/}). It integrates various catalogs and measurements of solar flares from over a dozen data sources \cite{Sadykov17}. IMDSF provides convenient search and quick-look data overviews. The Heliophysics Event Knowledgebase (HEK, \url{https://www.lmsal.com/heksearch/}) organizes observational data from several space instruments as a joint data hub that allows users to perform search and visualization by identified event class and observing characteristics \cite{Hurlburt12}. The Space Weather Database Of Notifications, Knowledge, Information (DONKI, \url{https://ccmc.gsfc.nasa.gov/donki/}) is an online search tool associated with high energy-release related events, such as flares, coronal mass ejections, solar energetic particles, geomagnetic storm events, etc. For each event, DONKI provides a brief report and a list of associated events and related models (if available). These examples show that data integration from various sources, as well as advanced search capabilities, stimulate and enhance the scientific outcomes from the data \cite<e.g.,>{Nita04,Milligan18}.

In this paper, we present the developed Radiation Data Portal, which accumulates radiation measurements at aviation altitudes and soft X-ray and proton flux measurements from GOES satellites. The Radiation Data Portal is an interactive web-based application for convenient search and visualization of in-flight radiation measurements and exploration of various properties related to the radiation environment. Section~\ref{section:backend} describes the database back-end structure and Application Programming Interface (API). The portal front-end search and visualization capabilities are described in Section~\ref{section:frontend}. An example of search and retrieval of ARMAS measurements during solar proton events (SPEs) is described in Section~\ref{section:usecase}, followed by a discussion in Section~\ref{section:discussion}.

\section{Radiation Data Portal Back-End}
\label{section:backend}

Development of the Radiation Data Portal requires efficient handling and storage of the incoming data. In this section we describe the Portal data handling environment. In particular, we describe a list of currently-integrated sources, their structure and organization in the database, and the capabilities of the portal Application Programming Interface (API) to request and retrieve data records.

\subsection{Data Sources}

The Radiation Data Portal currently integrates the following data sources:

\begin{itemize}
\item \textbf{ARMAS radiation measurements} augmented with the integrated properties of the flight and environment. A detailed description of  ARMAS measurements can be found at \citeA{Tobiska15,Tobiska16,Tobiska18}. In essence, the ARMAS project utilizes a micro-dosimeter integrated into a data processing and communication electronics package to measure and report the absorbed dose rates with a one-minute cadence. The doses are converted to the effective dose rates. Currently, the ARMAS data are publicly available from Space Environment Technologies (SET) as individual files for each flight (\url{https://spacewx.com/radiation-decision-aids/}). The files contain information about the general environment conditions during a flight, such as indexes of geomagnetic activity (Kp, Ap, G-level), physics-based modeled radiation dose rates from the climatological NAIRAS v1 model \cite{Mertens13}, and measurements of protons above 1.0\,MeV and electrons above 0.6\,keV from GOES.
\item \textbf{GOES Soft X-ray (SXR) radiation in the 0.5-4\,$\AA$ and 1-8\,$\AA$ channels}. The Geostationary Operational Environmental Satellites (GOES series) traditionally have onboard X-ray sensors \cite<XRS,>{Bornmann96}, and measure SXR fluxes in two channels. The Radiation Data Portal currently utilizes calibrated 1-min averaged GOES fluxes available from the National Oceanic and Atmospheric Administration National Centers for Environmental Information archive (NOAA NCEI, \url{https://satdat.ngdc.noaa.gov/}). Each measurement during the flight is connected to the nearest-time SXR measurement.
\item \textbf{Integrated GOES proton flux measurements.} These measurements are taken by two Energetic Proton, Electron, and Alpha Detectors \cite<EPEADs,>{Bruno17} onboard the GOES~15 satellite. These detectors are pointed in the east and west directions. The Portal utilizes 5-min calibrated and integrated measurements of the proton fluxes above the following thresholds: 1\,Mev, 5\,MeV, 10\,MeV, 30\,MeV, 50\,MeV, 60\,MeV, and 100\,MeV. The measurements are averaged for the east and west EPEAD detectors. These data are available via the NOAA NCEI archive (\url{https://satdat.ngdc.noaa.gov/}). Each measurement during the ARMAS flight is connected to the closest-in-time GOES measurement.
\end{itemize}

The \remove{described} data are collected starting from 2013 and integrated into the MySQL database.

\subsection{MySQL database}

\begin{figure}
\includegraphics[width=1.0\linewidth]{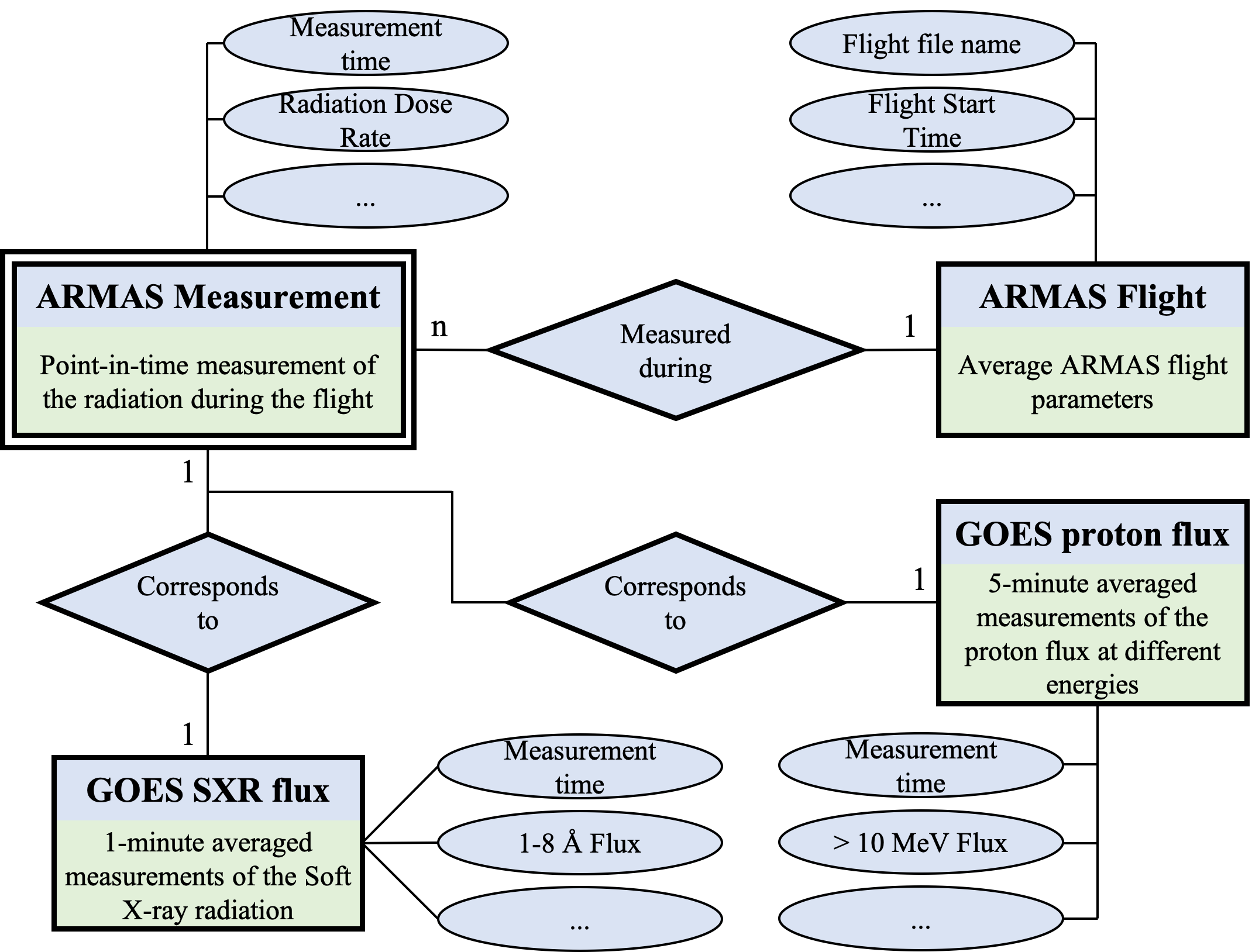}
\caption{The entity relationship diagram of the Radiation Data Portal.}
\label{figure:figure1}
\end{figure}

The described data are loaded into the MySQL relational database. The database structure, illustrated in Figure~\ref{figure:figure1}, represents an interaction between three entities: 1) integrated ARMAS flight properties, 2) measured time-series during a flight, and 3) GOES measurements. In the present time, the ARMAS flight measurements update is performed manually using the developed update scripts. The data update procedure includes loading of new in-flight measurement records into the database and checking for data calibration and format changes of new data sets provided by Space Environment Technologies (SET). In the case of significant changes to ARMAS data source files, the structure of the database is adjusted manually.

\subsection{Application Programming Interface (API)}

The API developed for the Radiation Data Portal at present handles the following two types of requests: 1) properties of ARMAS flights, and 2) measurements during a flight. Properties of ARMAS flights can be requested by a user-specified time range or a flight identifier, which corresponds to the file name in the original ARMAS archive. A request of measurement obtained during a specific ARMAS flight can be submitted by a flight identifier. These requests are implemented as HTTP GET requests. The web API documentation (the description of functions and related Python routines) are available from the Radiation Data Portal web page. As an example, a request for measurements during the ARMAS flight for September 7, 2017, is available via the following URL: \url{https://data.nas.nasa.gov/helio/portals/rdp/ARMAS_API/apiget_flightmeasurements_byID.php?flightID=ARMAS_dirIP_Report_56940_20170907013150_L1L4.txt}

\section{Radiation Data Portal Front-End. Web Application}
\label{section:frontend}

For convenient search and overviews of ARMAS and GOES data, we have developed a comprehensive front-end web application. It includes a search engine and a description of the project and data sources, as well as contact information. The Radiation Data Portal is currently deployed at NASA Advanced Supercomputing (NAS) facilities at \url{https://data.nas.nasa.gov/helio/portals/rdp/}. In case if any issues with the data or the portal, the users may contact the developers using the contact information provided in the paper or the database webpage.

\subsection{Data Search Filters and Visualization}

The search engine provides selection of flights with properties specified by the user. The search filters are organized into three categories.

\begin{itemize}
\item \textbf{Flight time and location properties.} This category of filter is based on temporal flight properties: dates when flights were performed and their duration and the location of flight routes such as longitudes, latitudes, altitudes, geomagnetic longitudes and latitudes, and L-shells (a measure in Earth's radii of how far out from Earth's center the local magnetic field lines go when they cross the Earth's magnetic equator).
\item \textbf{Environment characteristics.} The filters in this category allow a user to select flights according to the background properties of the environment where the flights were taken. It includes space weather and geomagnetic indexes (Kp, Ap, D, Dst, G-level), peak values of 1-8\,$\AA$ SXR flux, and peak fluxes of protons above 10\,MeV and above\,100 MeV during the flight.
\item \textbf{Dosimetry measurements.} This category includes total, averaged, and median dose measurements obtained during ARMAS flights, as well as measurements predicted by the NAIRAS~v1 model, and correlations between the model and measurements. Additional properties such as geomagnetic cutoff rigidity and median quality factor fall into this category.
\end{itemize}

The filters can be applied by selecting the range of parameters and/or by a drop-down menu. The search results are visualized in the form of dynamic histograms according to applied filters. The visualization of histograms are implemented using the Google Charts API.

\subsection{Data selection and per-flight visualizations}

Once the filters are set, the user can push the ``Search'' button and receive a list of the flights satisfying the query. The list includes the time and duration of each flight, filenames, and links to download the original ARMAS files. The full list with the average flight properties and dosimetric measurements are available for download in a comma-separated CSV format. 

\begin{figure}
\includegraphics[width=1.0\linewidth]{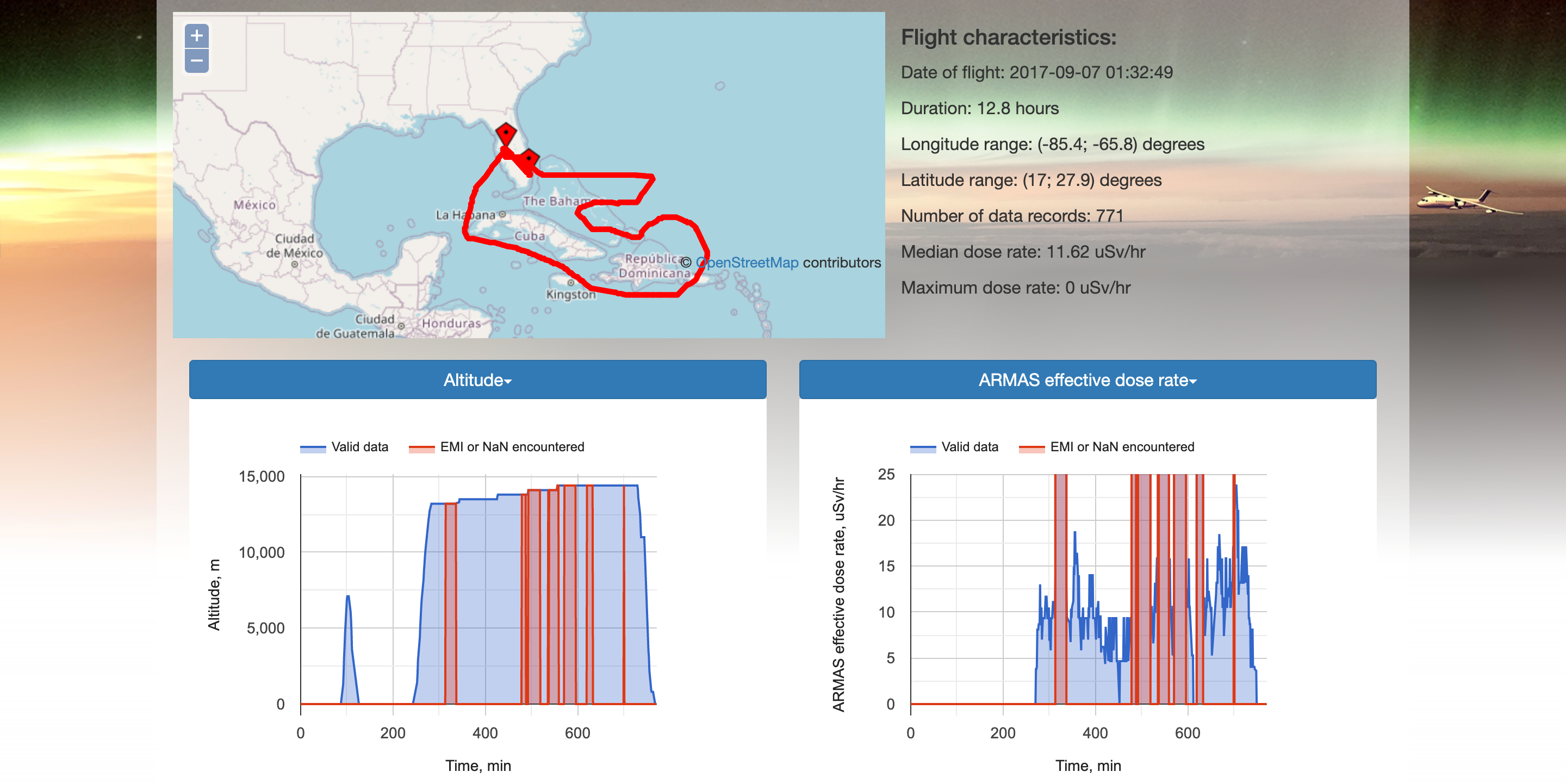}
\caption{Illustration of a quick-look visualization for the ARMAS flight on September 7, 2017.}
\label{figure:figure2}
\end{figure}

Each selected flight is supported by a quick-look visualization. An example quick-look visualization for the ARMAS flight on September 7, 2017 is presented in Figure~\ref{figure:figure2}. The visualization panel consists of four primary elements. The first element is an overview of a flight route on the map. The route mapping was implemented with the OpenLayers API. The second is a summary of the key properties of a flight. The third and fourth windows are charts presenting the flight parameters as functions of time. The line charts are implemented using the Google Charts API. The charts have the capability to select a flight property from a drop-down menu. The parameters as functions of time can also be downloaded in a comma-separated CSV format.

\section{Use Case Example: ARMAS Measurements in an Enhanced Solar Proton Environment}
\label{section:usecase}

To demonstrate how the Radiation Data Portal can enhance the scientific outcome from radiation measurements, we can select all flights occurring during enhanced fluxes (more than 10 particle flux units, pfu) of energetic protons above 10\,MeV. The peak energy threshold of 10\,MeV and proton flux exceeding 10\,pfu is used by the NOAA Space Weather Prediction Center (SWPC) as a definition for the ``proton event''. The results for six ARMAS flights are summarized in Table~\ref{table:table1}. A detailed study of flight parameters and measurements is based on a query from the Radiation Data Portal. The example Jupyter Notebook with the query and analysis of the results available at \url{https://github.com/vsadykov/RDP_example.}

\begin{figure}
\includegraphics[width=1.0\linewidth]{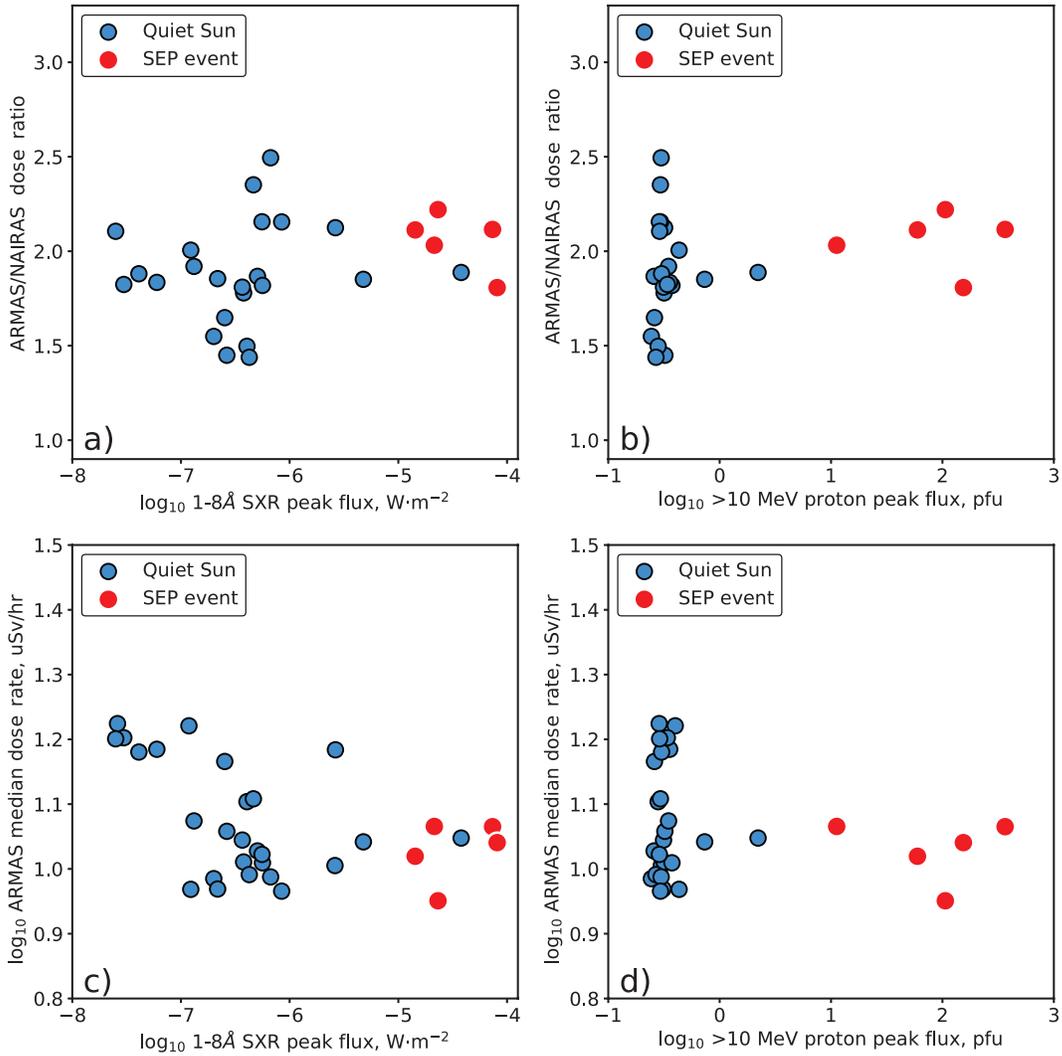}
\caption{Correlations between the total dose ratios of ARMAS to NAIRAS v1 to X-ray flux (panel a) and proton flux above 10\,MeV (b). Dependence of the ARMAS median effective dose rate from X-ray flux (panel c) and proton flux above 10\,MeV (d). Red circles correspond to flights when the peak value of $>$\,10 MeV proton flux exceeded 10\,pfu, and blue circles indicate the flights when the $>$\,10\,MeV proton flux was less than 10\,pfu.}
\label{figure:figure3}
\end{figure}

Figure~\ref{figure:figure3} illustrates the correlation between the total dose ratio of ARMAS measurements and the NAIRAS v1 model (hereafter, ARMAS/NAIRAS dose ratio) and soft X-ray (panel a) and proton (panel b) peak fluxes. The red points correspond to flights that occurred during solar proton events (SP-enhanced group of flights). The blue points represent flights with similar altitudes and route areas occurred when no solar proton events were observed (SP-quiet group of flights). These flights were performed from September 6, 2016 to September 6, 2018 at a longitude range of -95.5~----41.5 degrees and a latitude range of -3.0~---39.0 degrees. The median altitudes of these flights were in the range 13.1~---15.1 km. Panels c and d show the median values of the effective dose rates versus the SXR and proton peak fluxes. One of the flights (Flight 1 started at 2017-09-04 16:20:00) did not have a reported measurement of the median value due to long-lasting electromagnetic interference (EMI) periods and related spikes in the data. We excluded this flight from the plots.

To understand if the observed solar proton events are reflected in the ARMAS measurements we perform two statistical tests: 1) Student's t-test examines the statistically significant difference between the mean values of two distributions assuming equal variance, and 2) Welch's t-test also checks the statistically significant difference between the mean values of two distributions without the assumption of equality of the variances. These tests were applied to the ARMAS/NAIRAS dose ratio and median effective dose rates of SP-enhanced and SP-quiet groups of the ARMAS flights. For each test, we calculated the p-value, which represents the probability to incorrectly accept the hypothesis that the mean values of two distributions are different. We found that the p-values for the tests never dropped below the 0.10 value, indicating no statistically-significant differences between these two groups of flights.

The absence of differences between the two groups of flights may have the following possible explanations. Although the fluxes of $>$\,10\,MeV protons were high (above 10\,pfu), the fluxes of $>$\,100\,MeV protons never exceeded 0.25\,pfu during the flights. Probably, the energy carried by most of the particles was not sufficient to cause radiation disturbances in the Earth’s atmosphere at aviation altitudes. Interestingly, even during the SPE of September 10, 2019 with $>$50\,pfu peak flux of $>$100\,MeV protons \cite{Jiggens19}, the modeled solar energetic particle-related contribution to the dose rates at aviation altitudes was lower than the galactic cosmic ray-related contribution \cite<see Figure 3 of>{Kataoka18}. Another possible explanation is that the data-driven NAIRAS~v1 physics-based model is capturing much of what ARMAS is measuring since the ARMAS/NAIRAS dose ratio is relatively flat for these proton events. Given the fact that the number of flights in the SP-enhanced group was very low for the statistical analysis (only 5 flights with the determined ARMAS median dosage rate), we conclude that the provided explanations require more detailed study based on a larger statistics of flights.

\section{Discussion and Conclusion}
\label{section:discussion}

The Radiation Data Portal provides an interactive web-based application for convenient search and visualization of in-flight radiation measurements and exploration of various properties related to the radiation environment. At present it utilizes the radiation dose-related measurements obtained with the ARMAS module during over 750 flights and 2$\times$10$^5$ science quality data records obtained since 2013. The data are organized into a database and accompanied by related environment SXR and proton flux observations from the GOES satellites. The database is supported by the API that allows a user to interact and retrieve the data. The front-end side of the Portal represents the web application equipped with filters and visualizations to customize the query and explore the flight properties and measurements obtained during selected flights.

The presented example (Section~\ref{section:usecase}) of the use of ARMAS flight properties and GOES observations to study the effect of enhanced proton fluxes on the Earth's radiation environment at aviation altitudes demonstrates the scientific and practical potential of the Portal. In particular, the use case shows that there is no statistically significant difference between the median dose rates and ARMAS/NAIRAS dose ratios for the flights performed during enhanced solar proton fluxes (above 10\,pfu for $>$\,10\,MeV protons) and during time periods when these fluxes were low. In general, understanding the influence of solar energetic particles (SEPs) of different energies on the properties of the radiation environment at aviation altitudes is not possible without linking ARMAS data and SEP-related measurements of GOES, Advanced Composition Explorer (ACE), ground-based neutron monitors, etc.

The Radiation Data Portal represents a public-private partnership between NASA Ames Research Center and Space Environment Technologies aimed to enhance scientific knowledge of the radiation environment's behavior and practical application of available radiation measurements, models, and space environment observations. It is of high importance to establish, enrich, and maintain a dialogue between public organizations, academia, and private partners in order to make progress in the broad space weather enterprise.

The Radiation Data Portal will be expanded by including additional radiation measurement sources and data visualization capabilities and improving the search engine to include new data sets. In particular, we plan to include measurements of ion flux and magnetic field in-situ measurements from NASA's Advanced Composition Explorer (ACE), Geomagnetic activity data from NASA’s Coordinated Data Analysis Web (CDAWeb), ground-based neutron monitor data, and other space-based measurements of energetic particles. We envision that the Radiation Data Portal will enhance our knowledge about solar-terrestrial interactions and Space Weather and that it will grow into a comprehensive collaborative effort involving many sides of our diverse community.

\begin{sidewaystable}
\caption{General properties of ARMAS flights during the enhanced proton flux from the Sun. The flights are selected for a peak value of the flux above 10\,pfu for the protons above 10\,MeV.}
\label{table:table1}
\centering
\scriptsize
\begin{tabular}{|c|cccccc|}
\hline
\textbf{Flight parameter}	&	\textbf{Flight 1}	&	\textbf{Flight 2}	&	\textbf{Flight 3}	&	\textbf{Flight 4}	&	\textbf{Flight 5}	&	\textbf{Flight 6}	\\
\hline
\textbf{Start Time}		&	2017-09-04 16:20:00		&	2017-09-05 16:43:49		&	2017-09-06 16:32:10		&	2017-09-07 01:32:49		&	2017-09-08 04:34:00		&	2017-09-08 20:01:30		\\
\hline
\textbf{Duration, hr}	&	8.96	&	8.18	&	8.58	&	12.8	&	8.21	&	5.66	\\
\hline
\textbf{Longitude range, deg}	&	(-64.1; -51.5)	&	(-68.0; -56.7)	&	(-79.9; -59.5)	&	(-85.4; -65.8)	&	(-85.3; -71.2)	&	(-85.5; -73.3)	\\
\hline
\textbf{Latitude range, deg	}	&	(13.0; 28.5)	&	(13.0; 29.8)	&	(13.0; 26.4)	&	(17.0; 27.9)	&	(18.9; 29.0)	&	(19.4; 27.9)	\\
\hline
\textbf{Altitude, km}	&	14.1	&	14.0	&	14.1	&	14.1	&	14.1	&	14.4	\\
\hline
\textbf{ARMAS median dosage rate, uSv/hr}	&	-	&	8.93	&	10.46	&	11.62	&	10.98	&	11.63	\\
\hline
\textbf{ARMAS/NAIRAS dosage flight ratio}	&	-	&	2.22	&	2.11	&	2.11	&	1.81	&	2.03	\\
\hline
\textbf{$>$10\,MeV proton peak flux, pfu}	&	13.0	&	106.1	&	59.6	&	364.9	&	154.1	&	11.2	\\
\hline
\textbf{$>$100\,MeV proton peak flux, pfu}		&	0.10	&	0.09	&	0.25	&	0.12	&	0.06	&	0.07	\\
\hline
\textbf{1-8\,$\AA$ SXR peak flux, 10$^{-5}$W/m$^2$}	&	5.54	&	2.32	&	1.43	&	7.36	&	8.12	&	2.14	\\
\hline
\textbf{Kp index}	&	4	&	4	&	4	&	4	&	4	&	4	\\
\hline
\textbf{Ap index}	&	29	&	29	&	29	&	29	&	29	&	29	\\
\hline
\textbf{D index}	&	0	&	1	&	1	&	1	&	2	&	2	\\
\hline
\textbf{G-level}	&	G0	&	G0	&	G0	&	G0	&	G3	&	G2	\\
\hline
\end{tabular}
\end{sidewaystable}

\acknowledgments
We acknowledge NASA Ames Research Center for the use of computational resources. We thank the NAS team for help with the Radiation Data Portal deployment. The original ARMAS data are publicly available from the ARMAS links to the archive database at \url{https://spacewx.com/radiation-decision-aids/}. The GOES proton flux and soft X-ray data are publicly available from the National Oceanic and Atmospheric Administration National Centers for Environmental Information archive (NOAA NCEI, \url{https://satdat.ngdc.noaa.gov/}). ARMAS data and corresponding NAIRAS model data provided by SET under agreement with NASA LaRC and NAIRAS PI Chris Mertens. We thank the GOES team for the availability of high-quality scientific data. The work was supported by the NASA Radiation Portal Development Grant NNX12AD05A, NASA ARMAS Dual Monitor SBIR contract 80NSSC19C0194, and NASA LWS RADIAN Contract 80NSSC18K0187.

\bibliography{radiationportal}

\end{document}